# Compensated current injection circuit, theory and applications.

Giorgio Fontana

University of Trento, Department of Information and Communication Technology,
Via Sommarive 14, I-38050 POVO (TN), Italy

This paper presents a detailed description, analysis and example of practical application of a wide frequency band voltage-to-current converter. The converter is characterized by a combination of positive and negative feedback loops. This feature allows compensation for parasitic impedance connected in parallel with the useful load, which in turn keeps an excitation current flowing through the useful load independent of its impedance. The simplicity of the circuit and its good electrical properties are additional advantages of the scheme.

## I. INTRODUCTION

Excitation circuits are required for active sensors, i.e. sensors that do not directly produce an electric signal proportional to the measured physical quantity. Active sensors are characterized by the property that the measured physical quantity modulates an electrical parameter of the sensor.
Among them are: temperature sensors, pressure sensors, position sensors, acceleration sensors and many other [1].
The measurement of the useful electrical parameter of the sensor is usually performed by the combination of an excitation circuit and a readout circuit. The sensor can be excited by a current and a voltage is measured, alternatively the sensor can be excited by a voltage and a current is measured.
In addition to direct impedance measurement, many other more sophisticated techniques, which might involve also time and frequency, are indeed based on these two fundamental methods.
Current and voltage generators are therefore critical items in sensor electronics and their accuracy always appears as a design goal in the development of sensor subsystems.
Speaking of voltage sources, both AC and DC, we see that the overall accuracy is usually dependent on few elements of the excitation circuit, essentially voltage references and voltage dividing networks.
In voltage driven sensors, usually there is no technical difficulty that prevents the application of the exact excitation to the sensor as it appears on the voltage reference. A four-wire voltage feedback excitation circuit can readily compensate the voltage drop due to the connecting wires. Other disturbances such as magnetic induction, thermoelectric effects, photoelectric effects, can be shielded, easily compensated or filtered out.

Differently, for current sources an intrinsic difficulty may arise from the fact that even if the source can produce a precise AC or DC current, it is difficult to convey the current to the sensor without dispersion.

Similarly to the voltage excitation, the current in the vicinity of the sensor can be readily measured and controlled, but it is sometimes physically impossible to separate the current that is useful for the measurement from the current flowing in parasitic impedances.

Evidence of this behavior might come, for instance, from the front sensing capacitive displacement sensor. In this class of sensors, the measurement current flows in one side of the armature, the dispersion current flows in the other side of the same armature towards the environment. Bootstrapped shielding electrodes or guard rings can mitigate the effect but no exact compensation can be made without a properly designed circuit.

The possibility of controlling the exact current flowing in a device, compensating for parasitic effects, may also find application in actuator driving circuits. Current driven actuators are indeed affected by current partition between the varying actuator impedance and the parasitic impedances. The ability of compensating for parasitic currents may maximize the accuracy of the actuation.

This paper describes a simple circuit capable of generating an almost constant and controllable AC/DC excitation current in a varying transducer impedance, even if a known and invariant parasitic impedance is connected in parallel to the effective transducer impedance.

Analytical solutions are computed for the simplest case, whilst SPICE simulations are employed for the numerical analysis of a more realistic application.

Because of the inherent simplicity of the circuit and of its electrical components, SPICE simulations are indeed appropriate for demonstrating the validity of the technique [2], nonetheless a breadboard proof of concept circuit has been built, mainly for overall stability testing.

## II. THE COMPENSATED CURRENT INJECTION CIRCUIT

The Compensated Current Injection (CCI) circuit is shown in figure 1. The electric network is composed by the effective sensor (actuator) impedance $Z_S$, the global parasitic impedance $Z_P$, three additional impedances $Z_1$, $Z_2$, $Z_3$, an instrumentation differential amplifier with a gain of 1 and an operational amplifier with gain $A_o$.

In the analysis $Z_P$ represents the parallel of all the parasitic impedances of the sensor (actuator), wiring, amplifier input impedance and bias resistors. $Z_3$ represents the parallel of a discrete passive electric network, the amplifier input impedance and bias resistors.

The transfer function of the CCI circuit can be calculated with result:

$$I_S = V_{IN} \frac{Z_P(Z_2+Z_3)A_o}{Z_P Z_1[Z_2+Z_3(1+A_o)]+Z_S[(Z_P+Z_1)(Z_2+Z_3)+A_o(Z_1 Z_3-Z_P Z_2)]} = V_{IN} Y_T \qquad (1)$$

The expression has been written in a form that puts in evidence the open loop transfer function of the operational amplifier $A_o$ and the effective sensor (actuator) impedance $Z_S$.

By inspecting equation (1) a balance condition can be easily identified by choosing:

$$Z_1 Z_3 = Z_P Z_2 \qquad (2)$$

The balance condition can be satisfied by means of direct measurement or analytical prediction of the different impedances therein, except at least one, which must be imposed using equation 2. At least one of the impedances in equation 2 must be adjustable for fine tuning. The procedure of fine tuning is similar to bridge balancing, but the physical quantity measured during the balancing procedure is application dependent, this point will be discussed in section IV.
Provided that by some means the balance equation is satisfied we have:

$$I_S = V_{IN} \frac{Z_P(Z_P+Z_1)A_o}{Z_P Z_1[Z_1+Z_P(1+A_o)]+Z_S(Z_P+Z_1)^2} = V_{IN} Y_{TB} \qquad (3)$$

With the condition $A_o \to \infty$ equation (3) approximates to:

$$I_S = V_{IN} \frac{(Z_P+Z_1)}{Z_P Z_1} = V_{IN} Y_{T\infty} \qquad (4)$$

The transadmittance $Y_{T\infty}$ in equation (4) does not depend on $Z_S$ as required.
Equations (2) and (4) characterize the behavior of the CCI circuit for applications.
In a practical standpoint the balance condition (equation 2) can be satisfied with the highest possible accuracy, but $A_o$ cannot usually exhibit a modulus much higher than $10^6$, therefore equation 3 should be employed for the exact calculation of the transadmittance of a balanced CCI circuit. The complexity of equation 3 and the fact that $A_o$ is usually a transfer function suggests a simulation approach if an accurate analysis must be carried on a realistic CCI circuit.
The output impedance, which includes the parasitic impedance $Z_P$, can be computed by removing $Z_S$, then choosing $V_{IN}=0$ and calculating the voltage $V_{ZP}$ across $Z_P$ due to a now external current $-I_S$. The output impedance is therefore:

$$Z_O = \frac{V_{ZP}}{-I_S} = \frac{Z_P Z_1[Z_2+Z_3(1+A_o)]}{(Z_1+Z_P)(Z_2+Z_3)+A_o(Z_1 Z_3 - Z_P Z_2)} \qquad (5)$$

The following output impedance as seen by the effective transducer impedance $Z_S$, characterizes a balanced circuit:

$$Z_{OB} = \frac{Z_P Z_1}{(Z_1+Z_P)} + A_o Z_1 \left(\frac{Z_P}{Z_1+Z_P}\right)^2 \qquad (6)$$

We can recognize in equation (6) the zero gain ($A_o=0$) output impedance as the impedance of the parallel of $Z_P$ and $Z_1$. Increasing $A_o$ can increase the output impedance.
With the realistic assumption of $Z_P = Z_1$ we have:

$$Z_{OB} = \frac{Z_1}{2}\left(1+\frac{A_o}{2}\right) \qquad (7)$$

Which shows a simple to remember relation between the output impedance $Z_{OB}$, $Z_1$ and $A_o$.

As a final remark, it must be stressed that the CCI circuit is characterized by a positive and a negative feedback through separate electric networks. In a balanced circuit, stability is ensured by a finite load impedance $Z_S$. An open circuit load, associated to overcompensation might push the circuit to the limit of an intrinsic instability.

### III. NOISE ANALYSIS

There are numerous possible noise source in the CCI circuit, and some of them are not uniquely defined in the transfer function analysis because impedances $Z_i$ are somehow arbitrary. A complete noise analysis can be easily performed after the definition of the circuit impedances, and the analysis could be performed either analytically or numerically with simulation softwares like SPICE.

The exact expression of the output noise of the circuit is here calculated assuming impedances $Z_i$ pure reactances. The presented analysis is adequate for the application example of section IV.

We first analyze the effect of the voltage noise of the amplifiers.

By simple inspection and taking into account that the instrumentation amplifier has a gain of one, the output current noise density due to the voltage noise of the amplifiers is easily expressed in terms of the CCI circuit transfer function:

$$W_{Ien} = |Y_T|^2 (W_{enIns} + W_{enOPAMP}) \qquad (8)$$

where $W_{Ien}$ is the current noise spectral density in impedance $Z_S$ due to the voltage noise of the amplifiers, $Y_T$ is the CCI circuit transfer function, $W_{enIns}$ is the voltage noise spectral density of the instrumentation amplifier and, $W_{enOPAMP}$ is the voltage noise spectral density of the operational amplifier.

The effects of the current noise generators in the istrumentation amplifier are analyzed by using the expression of the impedances as seen by the inputs of the amplifier.

Input– sees the impedance $Z_- = Z_O // Z_S$, which can be easily computed.

For a balanced CCI the impedance $Z_-$ is:

$$Z_- = \frac{Z_1 Z_P Z_S [Z_1 + Z_P (1 + A_o)]}{Z_1 Z_P (Z_1 + Z_P) + Z_S (Z_1 + Z_P)^2 + A_o Z_1 Z_P^2} \qquad (9)$$

Using equation (9), taking the limit for $A_o \to \infty$ and under a balance condition the impedance seen by input – is therefore $Z_S$. This is the worst case for a balanced circuit, in fact with a lower gain the impedance is lower, less current noise is sent to $Z_S$ and less voltage noise can be computed as equivalent noise transferred to the input.

Input + sees the impedance:

$$Z_+ = \frac{Z_2 Z_3 [Z_1 Z_P + Z_S (Z_1 + Z_P (1 - A_o))]}{(Z_2 + Z_3)(Z_P Z_S + Z_1 (Z_P + Z_S)) + A_o (Z_3 Z_1 (Z_S + Z_P) - Z_2 Z_P Z_S)} \qquad (10)$$

It is no longer possible to apply a full balance equation, but a more restrictive balance equation could be applied.

Choosing:

$Z_1 = C Z_2$ and $Z_P = C Z_3$, $\qquad (11)$

with C a complex constant, the result is:

$$Z_{+B} = \frac{1}{C} \frac{Z_1 Z_P [Z_1 Z_P + Z_S(Z_1 + Z_P(1-A_o))]}{A_o Z_1 Z_P^2 + (Z_1 + Z_P)(Z_P Z_1) + Z_S(Z_1 + Z_P)^2} \quad (12)$$

The limit for $Z_{+B}$ for $A_o \to \infty$ is therefore $-Z_S/C$.

With a noise current injected in the non inverting input of the instrumentation amplifier a noise voltage appears, and the same voltage appears also on its inverting input, being the differential loop gain infinite.

Assuming negligible correlation among the different processes, the current noise density in impedance $Z_S$ due to the amplifiers and with $A_o \to \infty$ can be written as:

$$W_{inZs} = \left|\frac{(Z_P + Z_1)}{Z_P Z_1}\right|^2 (W_{enIns} + W_{enOPAMP}) + W_{inIns} + W_{inIns}\left|\frac{1}{C}\right|^2 \quad (13)$$

where $W_{inIns}$ is the current noise spectral density of the instrumentation amplifier with a gain of one. The current noise of the operational amplifier has been neglected because the impedance of the driving network is low, making the effect of the noise current negligible.

The current noise spectral densities named $W_{inIns}$ in equation 13 may also include, as an additional term, the thermal noise of bias resistors. Bias resistors can be easily incorporated in $Z_P$ and $Z_3$ for the calculation of the transadmittance.

Under the already discussed limitations of the analysis, equations 11, 13 are generally useful guidelines for choosing impedance $Z_1$ and the constant $C$, which in turn determines $Z_2$ and $Z_3$.

## IV. APPLICATION EXAMPLE: THE PLANE CAPACITOR FORCE ACTUATOR

The proposed application is the plane capacitor force actuator with parallel electrodes [3], driven by the CCI circuit.

Using sine AC current $I_S$ with frequency $f=\omega/2\pi$, the average attractive force between the two electrodes with area $S$ of the ideal plane capacitor having capacitance $C_S = \varepsilon S/x$ is:

$$F = \frac{\partial E}{\partial x} = \frac{\partial}{\partial x}\frac{C_S V^2}{2} = \frac{\partial}{\partial x}\frac{C_S}{2}\left(\frac{I_S}{\omega C_S}\right)^2 = \frac{\partial}{\partial x}\frac{1}{2 C_S}\left(\frac{I_S}{\omega}\right)^2 = \frac{\partial}{\partial x}\frac{x}{2\varepsilon S}\left(\frac{I_S}{\omega}\right)^2 = \frac{I_S^2}{2\varepsilon S \omega^2} \quad (14)$$

where the electrical quantities $V$, $I_S$ are expressed in RMS values and, if $x$ is a function of time, $\omega$ has to be much higher than the highest spectral component in $x$.

Equation 14 shows that if a controlled current drives this class of force actuators, the applied force does not depend on the distance $x$. An additional benefit is the lack of electrostatically induced stiffness $\partial F/\partial x$.

Figure 2 represents a CCI circuit connected to a plane capacitor actuator, characterized by a thin wire flexible connection. Both rear faces of the plane capacitor actuator, with active capacitance $C_S$, suffer of parasitic coupling to the environment. For the presented example only the right plate is affected by an effective capacitive parasitic coupling $C_P$ to ground potential. $C_P$ must be measured with the highest accuracy and must be time invariant for the proper operation of the circuit, the balancing procedure is described below.

An accurate measurement of the capacitance of $C_P$ is required for the design of the CCI network and for the determination of $I_S$ as shown in equation 16.

Being $C_P$ and $C_S$ connected in parallel, the total capacitance can be measured twice, imposing two different values of the distance $x$, which geometrically determines $C_S$. Then $C_P$ can be easily calculated.

Choosing the restrictive balance condition of equation 11, with C=1 we have:

$$C_1 = C_2 \text{ and } C_p = C_3 \tag{15}$$

Equation 4 gives the effective transducer current:

$$I_S = V_{IN} \omega (C_P + C_1) \tag{16}$$

that does not depend on $C_S$ as required.

Being $A_o$ not infinite, the driving CCI circuit does not exactly satisfy equations 2 and 4, the current $I_s$ is not independent from $C_S = \varepsilon S/x$ and therefore, because of equation 14, such a system is characterized by $\partial F/\partial x \neq 0$. Nevertheless, for a balanced system, the higher $A_o$ is, the lower $|\partial F/\partial x|$ becomes.

For the exemplified application, the balancing could be performed by measuring the electrostatically induced stiffness, then $C3$ could be tuned in order to minimize the stiffness in absolute value.

Methods for measuring the stiffness are beyond the scope of this paper and can be found in reference 4.

An array of CCI circuits could be employed for the electrostatic actuation of single test mass, multiple electrode accelerometers with many degrees of freedom. This application is typical for space based instruments. Sometimes the test mass must be electrically insulated [5] and care should be taken to ensure that the summation of the currents entering the test mass is zero, the residual current flows between the test mass and the frame supporting the active electrodes. For this class of accelerometers the force acting on each active electrode and the test mass is well approximated by equation 14.

## V. RESULTS AND DISCUSSION

The application example has been studied with SPICE [6]. Figure 3 depicts the schematic of the simulated CCI circuit.

For the simulated example of figure 3, the parasitic capacitance $C_P$ is 30 pF. The sensor (actuator) capacitance $C_S$ is stepped from 1 to 5 pF by steps of 1 pF. The chosen values are compatible with those presented in references [3] and [4].

A generalization of the CCI equations is required for the analysis of this circuit. More precisely the restriction of K=1 for the instrumentation amplifier must be removed, because in the example of figure 3 the gain K of the instrumentation amplifier is not one but is expressed by a transfer function.

Using the properties of Laplace transformations, equations 3, 4 and 6 can be easily rewritten in order to include the new definition of K.

The transadmittance of the balanced circuit with finite gain $A_o$ is:

$$I_S = V_{IN} \frac{Z_P(Z_P+Z_1)A_o}{Z_P Z_1[Z_1+Z_P(1+K A_o)]+Z_S(Z_P+Z_1)^2} = V_{IN} Y_{TB} \qquad 17)$$

With the condition $A_o \rightarrow \infty$ equation 17 approximates to:

$$I_S = V_{IN} \frac{(Z_P+Z_1)}{K Z_P Z_1} = V_{IN} Y_{T\infty} \qquad 18)$$

The output impedance of a balanced circuit is:

$$Z_{OB} = \frac{Z_P Z_1}{(Z_1+Z_P)} + K A_o Z_1 \left(\frac{Z_P}{Z_1+Z_P}\right)^2 \qquad 19)$$

The CCI circuit depicted in figure 3 is characterized by the use of three OPAMPs to build a circuit configuration very similar to the well know instrumentation amplifier configuration. Resistors R3 and R4=R5 have been employed to acquire useful gain from U1 and U2. It follows that K becomes the closed loop transfer function of a non-inverting amplifier made with an OPAMP [7].
K contributes to the loop gain and improves the ability of the circuit to keep the current independent from the changes of the capacitance of $C_S$.
Two equal resistive dividing networks (R6, R8 and R7, R9) reduce the loop gain in the positive and negative feedback path by a half. This additional network is required because resistors R7 and R9 permit also the addition of the input voltage with the same weighting than those presented to the outputs of U1 and U2. Therefore, in this implementation, the open loop gain $A_o$ is the open loop gain of a LF411 divided by 2.
In addition, a high pass filter made with C6 and R11 has been employed to stabilize the working point of U3 and a more realistic bias resistance of 10 MΩ is employed for R1 and R2.
A RC network composed by C7 and R10 has been introduced to properly shape the loop transfer function for improving the stability margin.
The circuit does not require a global DC feedback loop for controlling the offset of the operational amplifiers. In fact the first stage composed by U1 and U2 has a DC gain of 41, this voltage gain applies to the DC offset of U1 and U2. The second stage has unitary DC gain because of capacitor C6. If required, an additional RC high pass filter with RC values equal to the couple R11 C6 could be connected to the non-inverting input of U3 to minimize the overall offset voltage.
Using these values, SPICE simulation yields the transfer function presented in figure 4. In the simulations the input voltage is 1 V.
Between 100 kHz and 1 MHz the loop gain is insufficient because of the dominant pole compensation included in the chosen operational amplifiers, therefore this implementation of the CCI circuit is not able to effectively control the current at high frequency. Between 100 Hz and 10 kHz the behavior of the circuit appears acceptable. For frequencies below 10 Hz the network composed by C6 and R11 reduces the loop gain, and again the circuit is not able to control the current as requested in order to reduce the injection of low frequency noise.
Transient simulations have been carried out to discover any possible anomaly in the behavior of this specific implementation of the CCI circuit. None has been found, except for a voltage noise peak in the 10Hz to 100 Hz frequency region at the output of U3. The problem can be corrected with a couple of 10 MΩ resistor connected in parallel to C1 and C2 respectively. This compensation keeps the CCI circuit balanced and causes minor modification to the transadmittance at 1 kHz.

Figure 5 shows the transient response of the CCI circuit depicted by figure 3 to a sine signal with amplitude of 1 V and frequency of 1 kHz. This is a frequency within the region of acceptable behavior as shown in figure 4. The five curves are almost undistinguishable at this scale.

Figure 6 reports an enlargement of the transient response. With effective transducer capacitance increased from 1 to 5 pF, the current increases from 9.45 to 9.50 nA. These values must be compared to 9.19 nA as predicted by equation 18 at 1 kHz.

The band-pass behavior of the application example is not a limitation, because the useful properties of current control expressed by equation 14 can be exploited by using a sine signal at a single frequency.

The current noise density in $C_S$ =4 pF has been computed with SPICE. The result, expressed in A/√Hz is shown in figure 7. The current noise dependence on the constant C, which appears in equation 13, is also confirmed by SPICE analysis. Solid curve refers to $C2=C3$=30 pF, broken curve refers to $C2=C3$=60 pF. To keep invariant the time constants and satisfy the balance equation R2 is reduced to 5 MΩ if $C2=C3$=60 pF.

The simulation has been made with temperature equal to 27.0 C. Most of the noise is due to the thermal noise of R1 and R2 (about 41 fA/√Hz each for R1=R2=10 MΩ).

It must be observed that at low frequency the output impedance of this implementation of the CCI circuit is not very high because the loop gain is reduced by network C6, R11. At low frequency the current noise is therefore efficiently shunted by $C_S$, $C_P$ and R1 itself.

The CCI circuit of figure 3 has been wired up on a breadboard using standard capacitors for $C_S$, $C_P$, $C1$ and $C2$. The nominal capacitance of 33 pF has been chosen for $C_P$, $C1$ and $C2$. $C3$ is a 60 pF variable capacitor. Resistor R9 has been reduced to account for the output resistance of the signal generator. Resistors with a nominal tolerance of 1% have been employed.

A simple balancing procedure has been devised for the breadboard test circuit. First $C_S$ has been removed from the circuit, then $C3$ has been tuned in order to put the CCI circuit at the limit of the stability region, then $C_S$ has been again connected to the circuit. The current in $C_S$ has been computed by measuring the voltage output of U3 and using the known values of $C_S$, $C_P$ and $C1$.

After balancing, the circuit has proven to be unconditionally stable for $C_S$ higher than 1 pF.

Here the term unconditionally refers to the fact that the simple presence of the hands of the operator does not change the stray capacitance to the limit of inducing instability. A careful electrostatic shielding can avoid the manifestation of this effect.

Removing capacitor $C_S$ the circuit tends to oscillate at a frequency of about 1.5 kHz, which depends on stray capacitances. The reader should be aware of the fact that oscillations are due to positive feedback. As already stated with $C_S$ =0 the overall loop gain should be exactly zero, if with a slight detuning the loop gain becomes positive, then the CCI circuit oscillates at the frequency where the loop gain is maximum and the phase shift is zero.

The circuit has been tested at the operating frequency of 1 kHz, with supply voltages of ±15 V and input voltage of 1 V.

With $C_S$ =1 pF, the voltage at the output of U3 was 3.41 V and an injected current of 10.4 nA has been computed, with $C_S$ =5.7 pF, the voltage at the output of U3 was 0.64 V and an injected current of 10.5 nA has been computed. The property of the CCI circuit to keep constant the current in $C_S$ with $C_S$ =1 pF and $C_S$ =5.7 pF has been experimentally verified with an estimated accuracy not better than 10 %, because of additional, layout dependent, stray capacitances. In fact, the determination of the effective capacitances involved in the circuit from the measurements made on individual components is made slightly inaccurate by the capacitance of the wiring.

The maximum allowed output voltage of U3 and the current dividing properties of the network composed by $C_S$, $C_P$, and $C1$ determines the maximum available current in $C_S$.

At the operating frequency, the network composed by capacitors $C_S$, $C_P$, $C1$, $C2$ and $C3$ should also present to the output of U3 an impedance high enough to avoid current overload and saturation of U3 itself.

Both in breadboard testing and with transient simulations, saturation effects can be readily observed, in agreement with the argumentation discussed above.

[1] Jacob Fraden, *Handbook of Modern Sensors, Physics, Designs, and Applications.* (AIP Press, Springer, second edition, 1996).

[2] Vladimirescu, *The SPICE Book.* (New York, John Wiley, 1994).

[3] P. Touboul, B. Foulon, E. Willemenot, Acta Astronautica. **45**, 605 (1999).

[4] Towards low-temperature electrostatic accelerometry, Laurent Lafargue, Manuel Rodrigues, and Pierre Touboul. Rev. Sci. Instrum. **73,** 196 (2002).

[5] Y. Jafry and T.J. Sumner, Class. Quantum Grav. **14**, 1567 (1997).

[6] PSPICE 9.1 student edition, freely available at the Cadence www server:
   http://pcb.cadence.com/Product/Simulation/PSpice/eval.asp

[7] see, for instance, A. S. Sedra and K. C. Smith, *Microelectronic circuits* (4$^{th}$ edition, Oxford University Press, 1998).

Figure 1. The Compensated Current Injection (CCI) circuit.

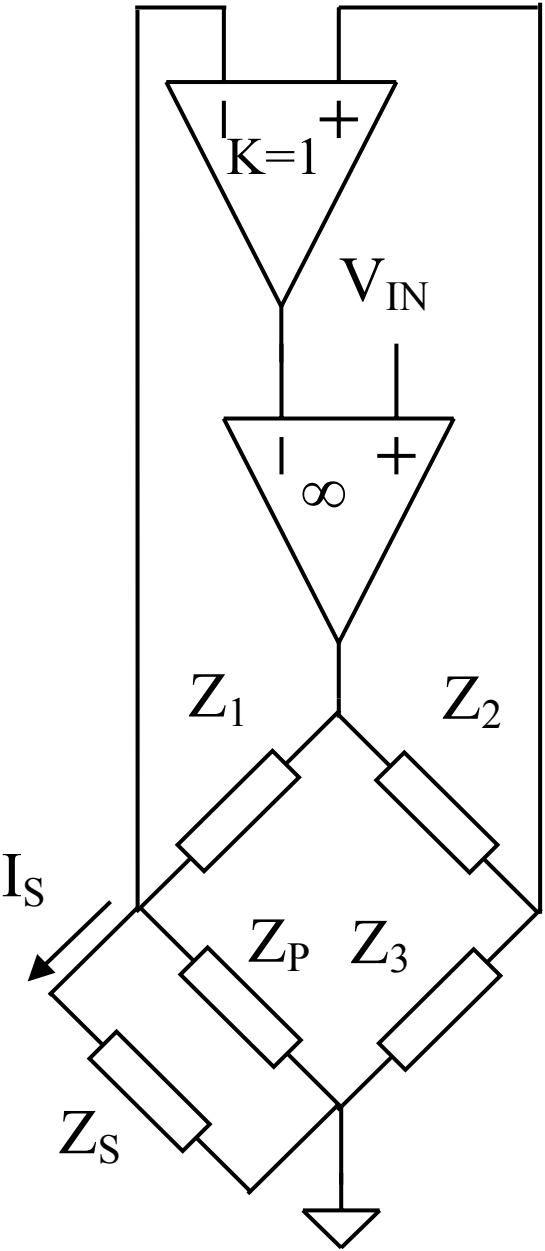

Figure 2. The CCI circuit connected to a plane capacitor electrostatic force actuator.

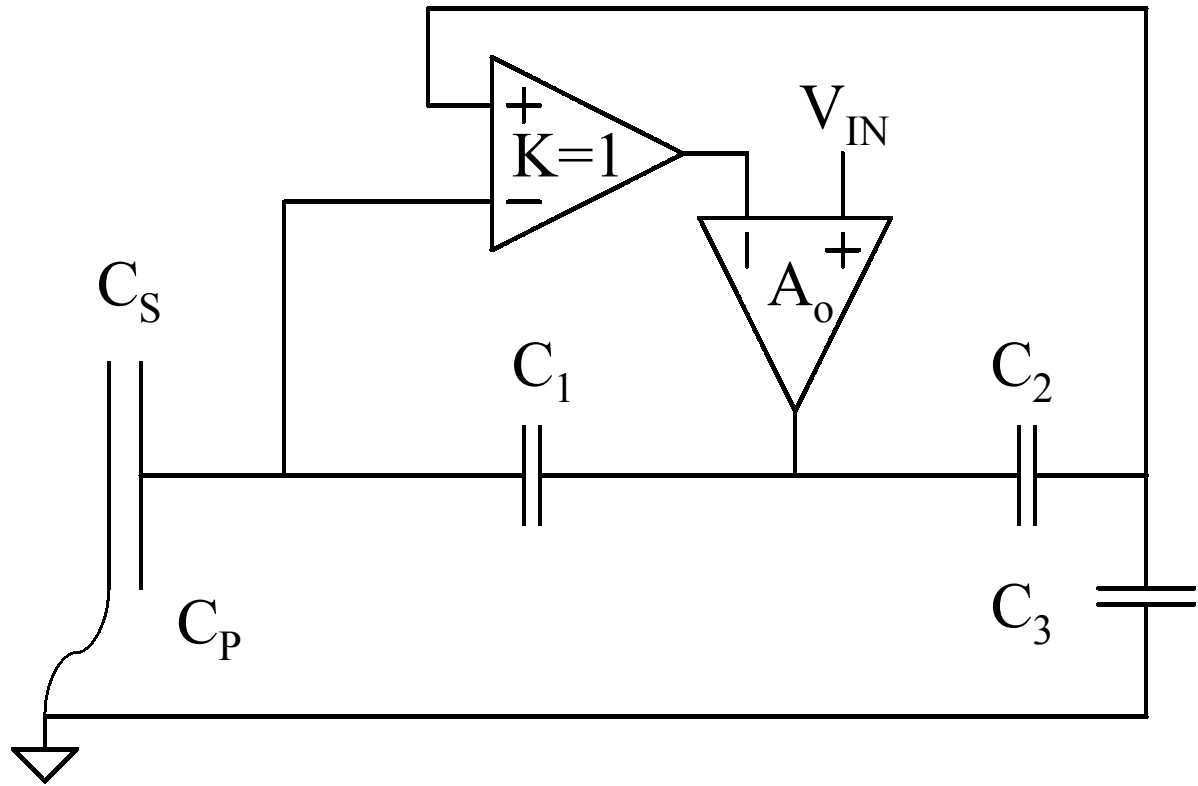

Figure 3. Schematic of the CCI circuit with real OPAMPs and capacitive load.

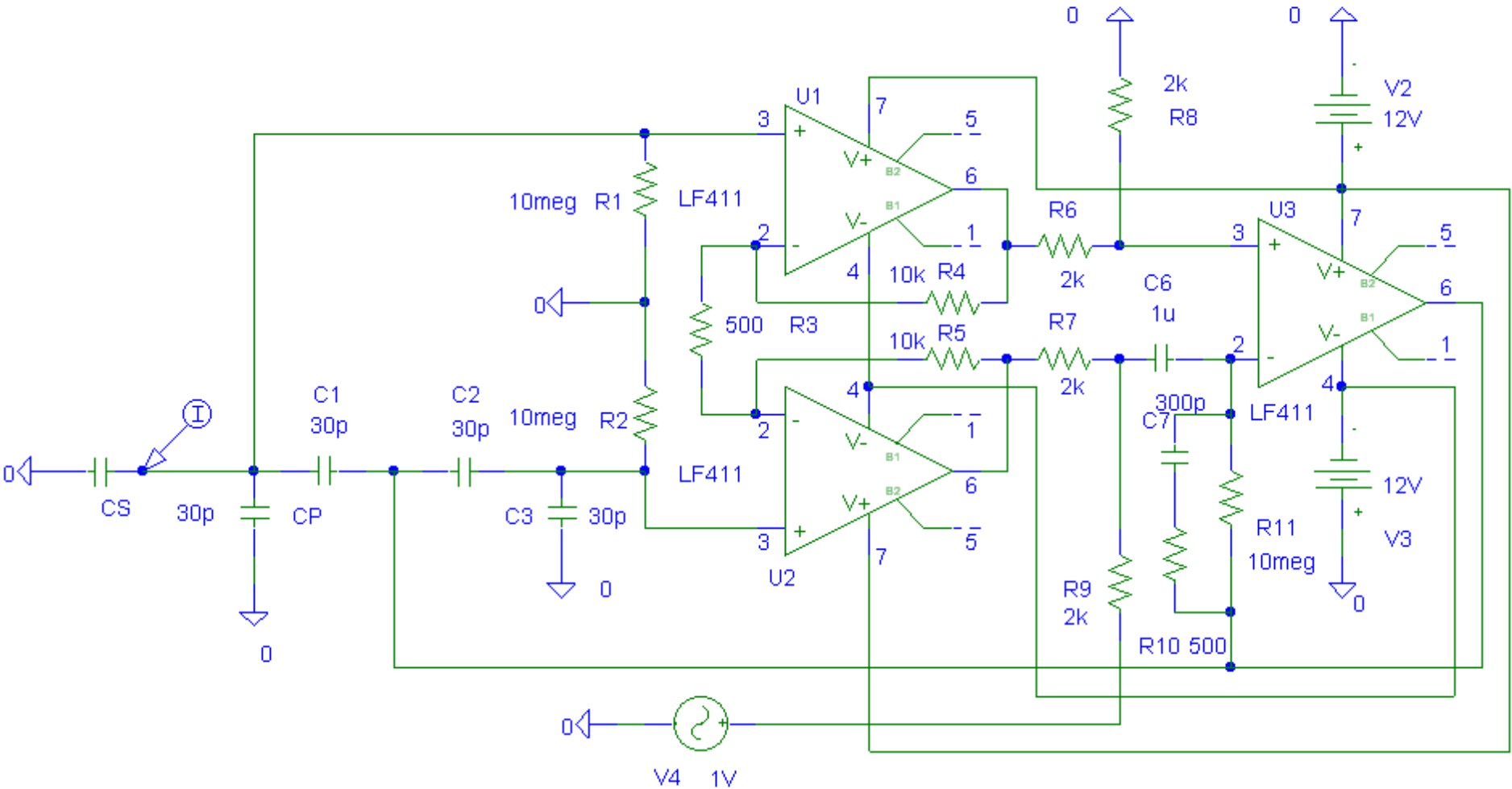

Figure 4. Simulated output current as a function of frequency with stepped load capacitance. Lower curve refers to $C_S = 1$ pF, upper curve refers to $C_S = 5$ pF, the step is 1 pF. Input voltage is 1 V.

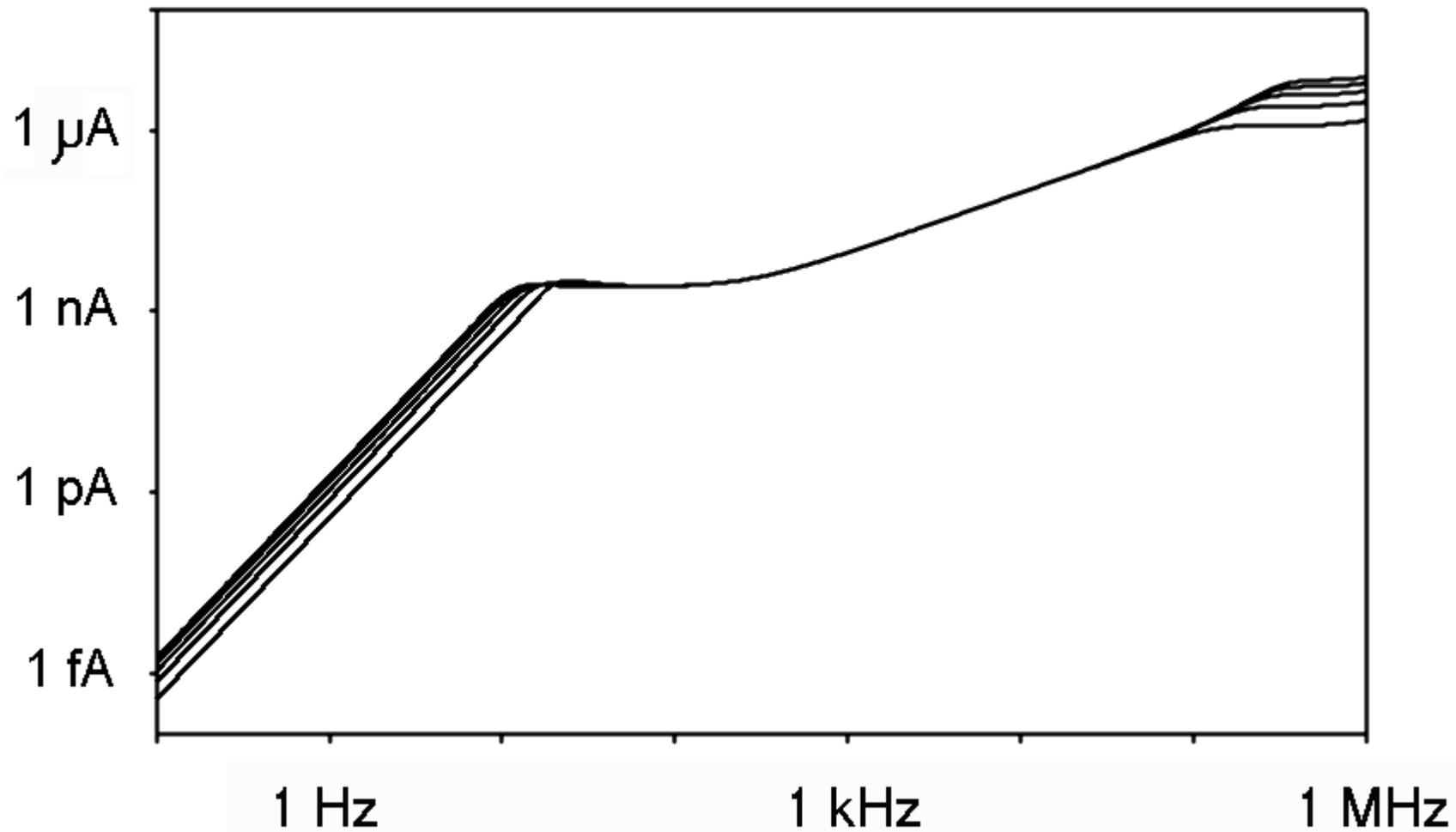

Figure 5. The simulated response to a sine signal with a frequency of 1 kHz. Input voltage is 1 V. The five output curves appear superimposed.

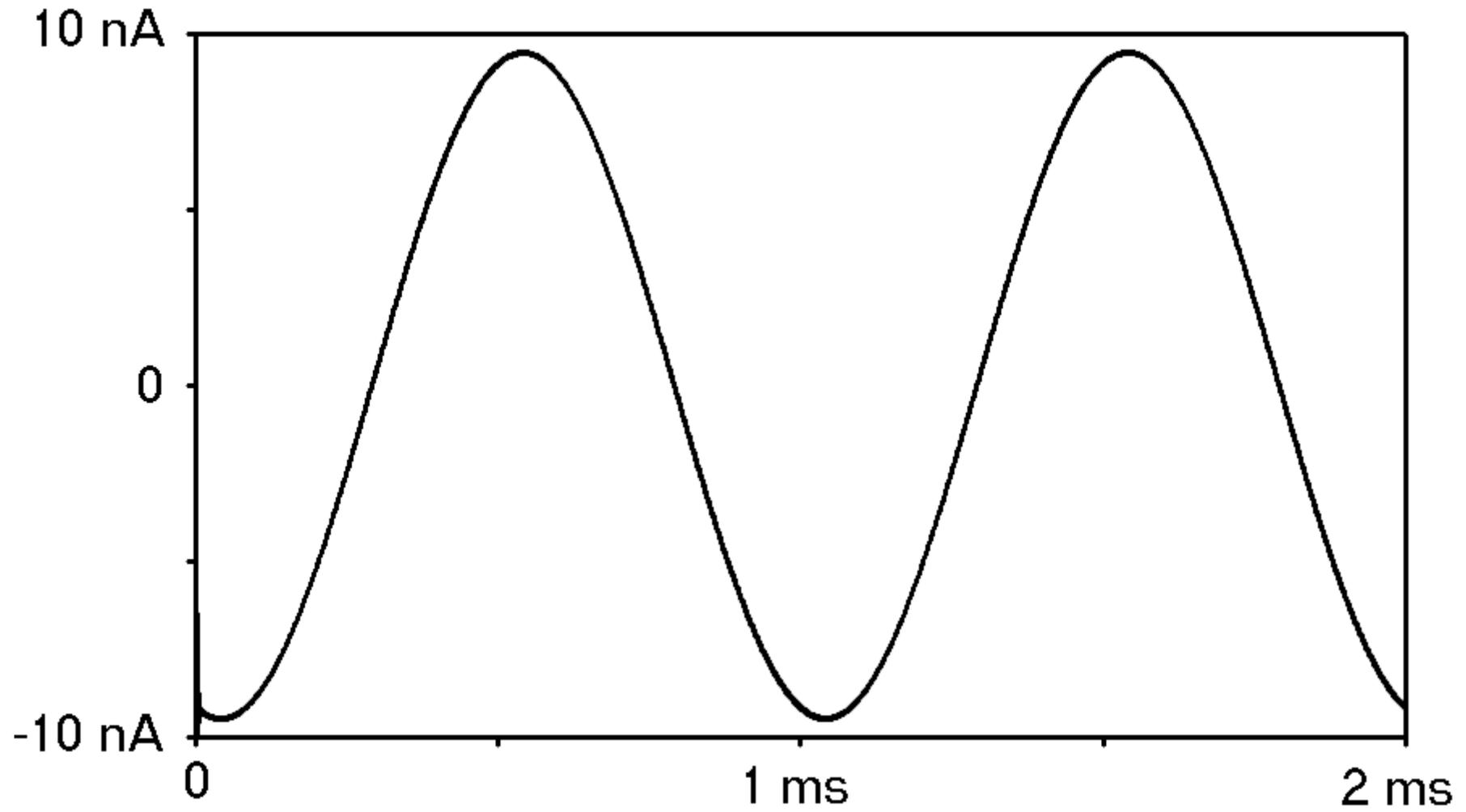

Figure 6. Enlargement of the simulated response to a sine signal with a frequency of 1 kHz. Lower curve refers to $C_S$ =1 pF, upper curve refers to $C_S$ =5 pF, the step is 1 pF. Sine signals are enlarged near their maximum values. Input voltage is 1 V.

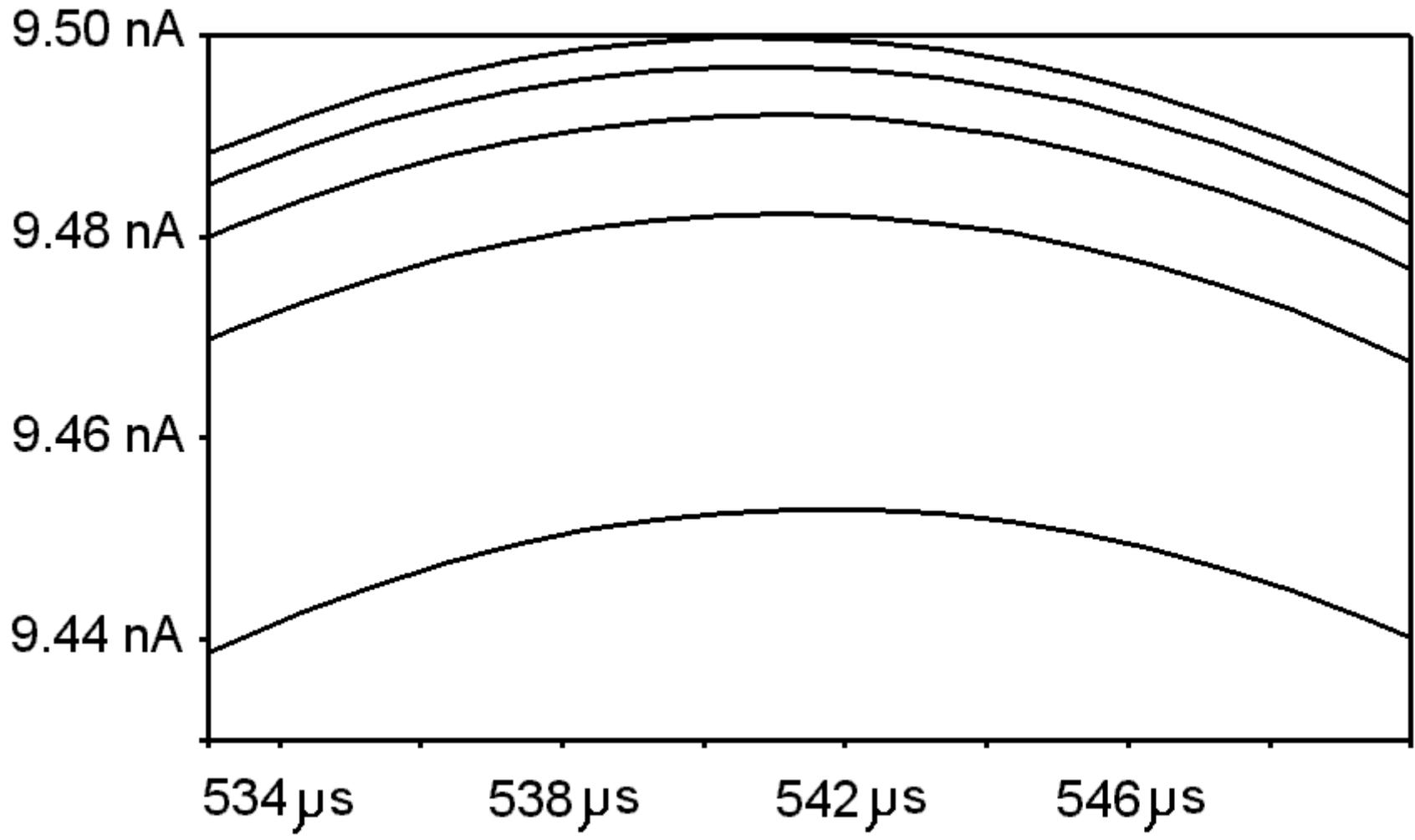

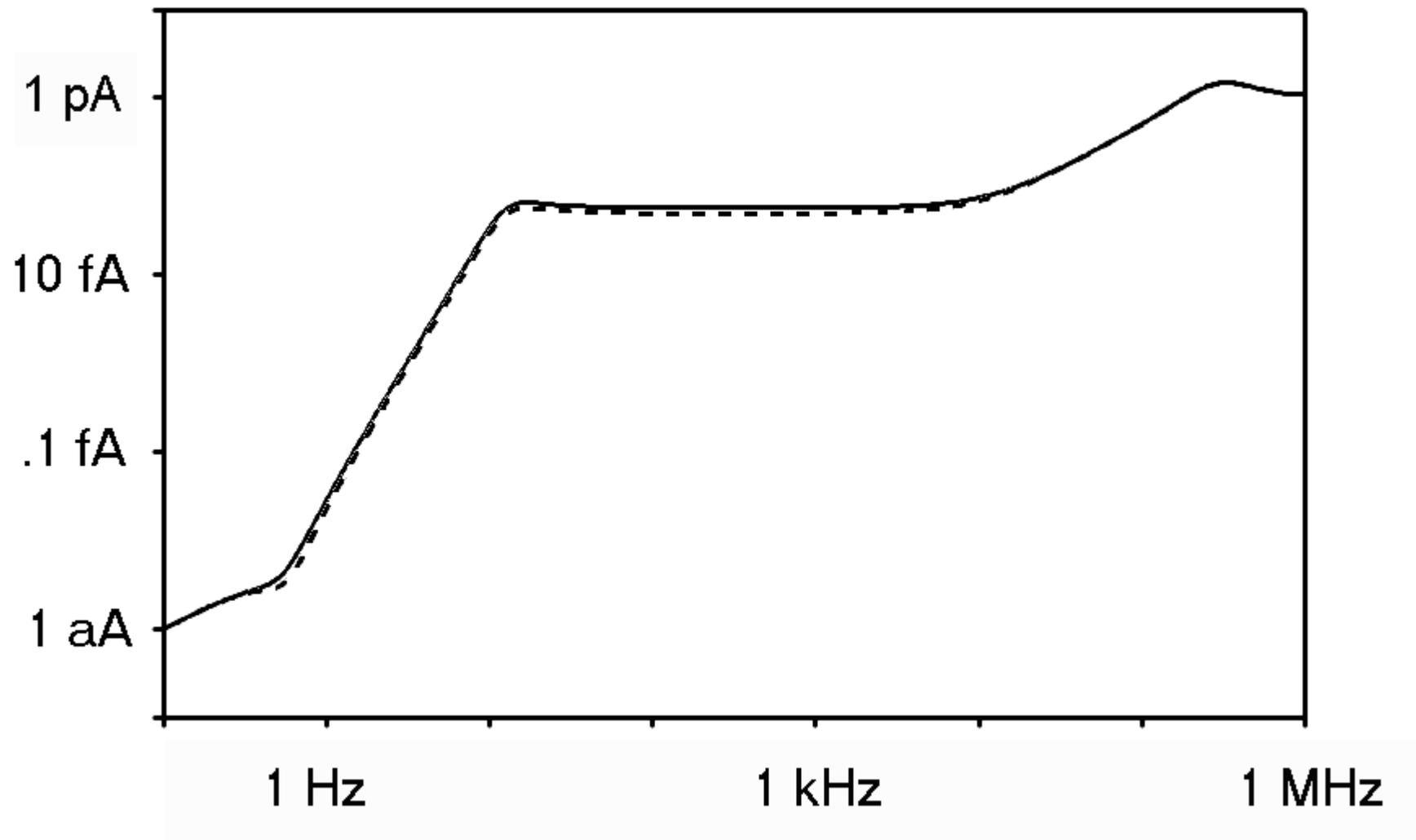

Figure 7. The current noise density in $C_S$ =4 pF computed with SPICE. The result is expressed in A/√Hz. Temperature is set to 27 C. Solid curve refers to *C2=C3*=30 pF, broken curve refers to *C2=C3*=60 pF.